# Does GLPT2 Offer Any Actual Benefit Over Conventional HF-MP2 In the Context of Double-Hybrid Density Functionals?

Golokesh Santra[1] and Jan M.L. Martin[1, a)]

[1]*Dept. of Molecular Chemistry and Materials Science, Weizmann Institute of Science, 7610001 Reḥovot, Israel*
[a)] Corresponding author: gershom@weizmann.ac.il

**Abstract.** While the inclusion of the nonlocal correlation in fifth rung "double hybrid" functionals is definitely beneficial, one might rightfully ask whether its evaluation in the basis of Kohn-Sham (KS) orbitals has additional value compared to the use of Hartree-Fock reference orbitals (in a type of multilevel scheme). We have investigated this question for a very large and chemically diverse dataset, GMTKN55. We conclude that KS reference orbitals are undoubtedly beneficial, but the benefit is not as large as one might intuitively expect.

## INTRODUCTION

The concept of double-hybrid density functional theory (DFT) appears to have arisen independently from two different sets of considerations.

From one side, Truhlar and coworkers[1,2] in 2004 proposed the MC3MPW and MC3BB "doubly hybrid" [sic] DFT approach, which are actually mixed WFT-DFT composite schemes that attempt to pair (a) the sound short-range performance of semilocal DFT correlation functionals, with (b) the good performance of second order Møller-Plesset perturbation theory (MP2) for dynamical long-range correlation. The late Angyán et al.[3] in pursuit of accurate van der Waals interactions, took this idea one step further in 2005, replacing a global linear combination by continuous range separation. (We note in passing that around the same time, several related approaches,[4–7] combining wave function theory and DFT were proposed primarily based on the "best of both worlds"[8] strategy.)

From the other side, Görling and Levy[9] proposed carrying out perturbation theory *in a basis of Kohn-Sham* (rather than Hartree-Fock) orbitals. Within Perdew's "Jacob's Ladder" of DFT, [10] this is a way to ascend from the second and third rungs (semilocal functionals) or the fourth rung (dependence on occupied orbitals) to the fifth rung (dependence on virtual orbitals). Grimme's B2PLYP functional[11] represents the first practical implementation of this concept, and is the first "double hybrid" functional (note their slightly different to have seen wide practical use. Here, KS orbitals are first generated with a hybrid DFT functional with damped semilocal correlation, and a nonlocal correlation term is added in by carrying out MP2 in a basis of the KS orbitals obtained. (The non-Brillouin single-excitation terms are neglected: T. van Voorhis, in 2008, communicated to the senior author that their inclusion had insignificant effects.)

In the wake of this work, a variety of empirical as well as "nonempirical" (also sometimes referred to as parameter-free) double hybrids have been proposed: for reviews, see e.g. Goerigk and Grimme,[12] Xu,[13] Adamo et al., [14] and the present authors.[15]

The excellent performance, for very large and chemically diverse thermochemical benchmarks like GMTKN55[16] (general main-group thermochemistry, kinetics, and noncovalent interactions, 55 problem subsets) and MGCDB84[17] (main-group chemistry database with about 5000 test cases divided into 84 subsets), of functionals like the Berkeley ωB97M(2),[18] Shanghai XYG7,[19] and Weizmann revDSD-PBEP86-D4[20] would seem to provide ample justification for the second approach. However, a fundamental question arises: does GLPT2 nonlocal correlation bring anything to the table that simple HF-MP2 does not? If no, then they would not have any significant advantage over the earlier composite WFT-DFT schemes. But if yes, how substantial is the improvement, and where might it stem from?

This is the question we seek to address in the present short note.

## COMPUTATIONAL METHODS

We have considered seven global double hybrids developed in our group for the present study: revDSD-PBEP86-D3BJ,[20] xDSD$_{75}$-PBEP86-D3BJ,[21] XYG8[*f1*]@B$_{20}$LYP,[22] XYG8[*f1*]@B$_{50}$LYP,[22] XYG8[*f1*]@B$_{60}$LYP,[22]

XYG8[*f1*]@B$_{70}$LYP[22] and XYG8[*f1*]@HF-LYP.[22] The dataset we employ is GMTKN55, which entails almost 2,500 species and 1,500 reaction energies: see Table 1 in Ref.[16] and associated text for a detailed description of the subsets and provenance of the reference values, obtained from coupled cluster ab initio calculations extrapolated to the complete basis set limit. WTMAD2 (weighted total mean absolute deviation, type two), as defined in the same paper,[16] has been used as our primary metric of choice for this entire study.

Q-Chem 5.4[23] was employed for the GLPT2 based revDSD-PBEP86-D3BJ, xDSD$_{75}$-PBEP86-D3BJ double hybrids and canonical MP2 correlation energies. All the electronic structure calculations involving XYG8[*f1*] functionals were performed using the MRCC2020[24] package. The def2-QZVPP[25] basis set was used throughout except for the seven anion-containing subsets – where the def2-QZVPPD[26] was employed instead – and the C60ISO and UPU23 subsets, where we settled for the def2-TZVPP[25] basis set to reduce computational costFor further computational and optimization details, see the methods section of Ref.[20–22] Unrestricted HF and KS, rather than their restricted open-shell counterparts, were used throughout for the open-shell species. Unlike Truhlar's "doubly hybrids", we have used the same basis set for both DFT and canonical MP2 correlation energies. We will substitute simple HF-MP2 for KS-PT2, reoptimize all the parameters, and compare performance.

Detailed statistics are available as Electronic Supporting Information at http://doi.org/10.34933/wis.000398.

## RESULTS AND DISCUSSION

Both for revDSD-PBEP86-D3BJ and xDSD$_{75}$-PBEP86-D3BJ, using KS orbitals instead of pure HF reference does seem to help (see Table 1). For the xDSD$_{75}$-PBEP86-D3BJ, WTMAD2 deteriorates from 2.15 kcal/mol to 3.27 kcal/mol if we use canonical MP2 correlation and reoptimize all linear parameters. Unsurprisingly, the two parameters that change the most are the same-spin and opposite-spin MP2 coefficients: the lower-lying virtual KS orbitals imply smaller coefficients to reach the same PT2 correlation energy. The coefficients for the empirical dispersion correction ($s_6$, the other three parameters $s_8$, $a_1$, and $a_2$ are kept constant throughout[21]) remain more or less the same. Among the five top-level subcategories, performance for barrier heights and large-molecule reaction energies suffers most (by 0.39 and 0.40 kcal/mol, respectively) when replacing GLPT2 with HF-MP2 correlation. The same is observed for revDSD-PBEP86-D3BJ (see Table 1).

Upon detailed inspection, it appears that much of the difference is caused by two of the total 55 subsets of GMTKN55: BH76 (the union of HTBH38[2] and NHTBH38[27]) and RSE43 (radical separation energies). Upon detailed inspection, we found that for the open-shell species in these subsets, $<S^2>$ values indicate severe spin contamination for HF-MP2, which is greatly mitigated in GLPT2 (it is well known, e.g.,[28] that DFT is much less prone to spin contamination). Upon eliminating the twelve reactions with the worst spin contamination (four from BH76, two from RSE43, one from RC21, and five from W4-11), WTMAD2 for xDSD$_{75}$-PBEP86-D3BJ only decreases from 2.15 to 2.08 kcal/mol using GLPT2 correlation, compared to a drop of 0.27 kcal/mol (3.27 to 3.00) for the HFMP2-based counterpart. Similarly, for the PT2-based and MP2-based revDSD-PBEP86-D3BJ variants, WTMAD2 values decrease from 2.37 and 3.12 kcal/mol, respectively, to 2.29 and 2.87 kcal/mol. Hence, removing critical spin contamination cases improves the performance of MP2-based functionals, but a significant gap still remains. (For perspective, it should be noted that 3.12 kcal/mol is in the same range as the 3.22 kcal/mol we obtain here for the ωB97M-V range-separated hybrid,[29] thus far the best global hybrid (rung four) functional reported. The reduction from the slightly higher value of 3.29 kcal/mol reported previously[15,20] is explained by the def2-QZVPPD basis set being used for BH76, which greatly reduces errors for the S$_N$2 Walden inversions.)

While the greater resilience to spin contamination of the KS orbitals is one part of the story, differences are also seen for subsets that are completely closed-shell, like 1,4-butanediol conformers (BUT14DIOL), where relative energetics are primarily driven by intramolecular hydrogen bonds. For subsets that are driven almost fully by London dispersion, like RG18 rare-gas clusters, ACONF alkane conformers, ADIM6 n-alkane dimers, it does not seem to matter whether you use regular HF-MP2 or GLPT2 correlation, as one might naïvely expect.

Differences in closed-shell subsets are instead seen where intermediate-distance interactions are significant, e.g., capped amino acid conformers AMINO20X4, peptide and sugar conformers PCONF21 and SCONF, or the BUT14DIOL set already mentioned. Interestingly, using GLPT2 correlations instead of regular HF-MP2 does more harm than good for reactions involving self-interaction error (SIE4x4) and tautomerization reactions (TAUT15).

Now, if we consider five XYG8[*f1*][22] type functionals and replace same-spin and opposite-spin GLPT2 correlation terms with corresponding canonical MP2 correlation components, WTMAD2 value increases gradually for the PT2-based variants with increasing HFx in the orbitals. However, for the MP2-based variants, interpolation suggests a minimum in WTMAD2 near 60% HF exchange in orbitals (i.e., XYG8[*f1*]@B$_{60}$LYP). The WTMAD2 gap between the PT2-based and MP2-based variants of XYG8[*f1*] decreases with increasing percentage of HFx used for orbital generation (see Table 1).

Comparing the five major subcategories: replacing PT2 correlation in XYG8[*f1*]@B$_{20}$LYP by MP2 correlation significantly degrades performance for all five. However, with the gradual increase of HFx in orbitals, small molecule thermochemistry, barrier heights, and large-molecule reaction energies become the driving force for the performance gap we see in total WTMAD2. In the case of XYG8[*f1*]@HF-LYP, a noticeable difference can only be seen for the large-molecule reaction energies.

**Table 1:** Total WTMAD2 (kcal/mol) for seven global double hybrid density functionals, as well as its decomposition into the five major subsets: THERMO=Thermochemistry of small and medium molecules; BARRIER=barrier heights; LARGE=reaction energies for large systems; CONF=conformer/intramolecular interactions; INTERMOL=intermolecular interactions

| Functionals | Nonlocal Corr. | WTMAD2 (kcal/mol) | THERMO | BARRIERS | LARGE | CONF | INTERMOL | NCI[b] |
|---|---|---|---|---|---|---|---|---|
| revDSD-PBEP86-D3BJ | PT2 | 2.372 | 0.515 | 0.269 | 0.555 | 0.463 | 0.570 | 1.033 |
|  | UMP2 | 3.117 | 0.563 | 0.536 | 0.829 | 0.596 | 0.593 | 1.189 |
|  | ROMP2 | 2.973 | 0.718 | 0.482 | 0.584 | 0.596 | 0.593 | 1.189 |
| xDSD75-PBEP86-D3BJ | PT2 | 2.145 | 0.494 | 0.250 | 0.487 | 0.432 | 0.481 | 0.914 |
|  | UMP2 | 3.273 | 0.562 | 0.641 | 0.878 | 0.619 | 0.573 | 1.191 |
|  | ROMP2 | 3.003 | 0.728 | 0.486 | 0.596 | 0.619 | 0.573 | 1.192 |
| XYG8[*f1*]@B20LYP | PT2 | 1.847 | 0.442 | 0.212 | 0.371 | 0.418 | 0.404 | 0.822 |
|  | UMP2 | 3.152 | 0.746 | 0.535 | 0.714 | 0.596 | 0.561 | 1.157 |
|  | ROMP2 | 3.098 | 0.855 | 0.562 | 0.523 | 0.596 | 0.561 | 1.158 |
| XYG8[*f1*]@B50LYP | PT2 | 1.935 | 0.451 | 0.227 | 0.377 | 0.432 | 0.448 | 0.880 |
|  | UMP2 | 2.982 | 0.674 | 0.497 | 0.719 | 0.538 | 0.556 | 1.094 |
|  | ROMP2 | 2.902 | 0.796 | 0.515 | 0.497 | 0.539 | 0.556 | 1.095 |
| XYG8[*f1*]@B60LYP | PT2 | 1.997 | 0.464 | 0.253 | 0.389 | 0.442 | 0.449 | 0.892 |
|  | UMP2 | 2.965 | 0.672 | 0.476 | 0.742 | 0.524 | 0.551 | 1.076 |
| XYG8[*f1*]@B70LYP | PT2 | 2.106 | 0.499 | 0.281 | 0.415 | 0.460 | 0.452 | 0.912 |
|  | UMP2 | 2.985 | 0.678 | 0.486 | 0.715 | 0.553 | 0.554 | 1.107 |
| XYG8[*f1*]@HF-LYP[a] | PT2 | 2.697 | 0.603 | 0.522 | 0.584 | 0.539 | 0.448 | 0.988 |
|  | UMP2 | 3.107 | 0.691 | 0.596 | 0.800 | 0.542 | 0.478 | 1.020 |
|  | ROMP2 | 2.895 | 0.798 | 0.526 | 0.504 | 0.583 | 0.484 | 1.067 |

[a] HF-LYP means 100% HF exchange, and for correlation 19% VWN5 and 81% LYP.
[b] NCI= Total noncovalent interactions (=CONF+INTERMOL)

Same as we observed for revDSD and xDSD functionals, the two most affected subsets are RSE43 and BH76. However, the performance gap gradually decreases with increasing percentage of exact exchange in the orbital generation step. Upon eliminating the twelve reactions with the severest spin contamination error, WTMAD2 remains practically unchanged for the PT2-based XYG8[*f1*]@B$_{20}$LYP, while it improves by 0.2 kcal/mol for its MP2-based counterpart. Interestingly, for the PT2-based XYG8[*f1*] functionals, WTMAD2 improves gradually with increasing %HFx in orbitals, but for the MP2-based cases, this improvement ranges from 0.2–0.3 kcal/mol throughout.

Upon further investigation, we found that performance for G21EA, AMINO20x4, HEAVY28, and to some extent RC21, PCONF21, and BHPERI degrades when we switch from GLPT2 to HF-MP2 correlation for the XYG8[*f1*] type functionals. Unlike what we observed for revDSD-PBEP86-D3BJ and xDSD$_{75}$-PBEP86-D3BJ, the SIE4x4 self-interaction subset benefits from using GLPT2 correlation instead of HF-MP2 correlation. For all those subsets, with increasing HF exchange in orbitals, performance gaps decrease gradually, and at 100% HFx, using MP2 correlation is marginally better than PT2 for the AMINO20x4 and BHPERI subsets.

The MP2-based XYG8[*f1*] functionals perform better than PT2-based variants for both the melatonin conformers (MCONF) and tautomerization reactions (TAUT15). For the rare gas clusters, PT2-based XYG8[*f1*]@B$_{20}$LYP is a better performer than its MP2-based counterpart, but for the XYG8[*f1*]@HF-LYP variants the trend is opposite. The choice of nonlocal correlation has little to no effect on the performance of XYG8[*f1*] family double hybrids for the ACONF alkane conformers, ADIM6 n-alkane dimers, conformers of inorganic systems (ICONF), HAL59 halogen bonds, IL16 cation-anion interactions, proton affinities (PA26), S66 noncovalent interactions, and WATER27 water clusters.

What happens when we use ROHF references for MP2? This causes significant improvements in, again, the BH76 and RSE43 subsets — somewhat offset by a deterioration in SIE4x4, where the stretching curves suffer from ROHF references. For xDSD75-PBEP86-D3BJ (and to a lesser extent revDSD-PBEP86-D3BJ), the gap with GLPT2 is significantly narrowed, yet the lion's share of the GLPT2 advantage remains. For XYG8@B20LYP and XYG8@B50LYP, the improvement when using ROMP2 is much smaller. For XYG8@HFLYP, a.k.a.

XYG8@B100LYP100%, however, the GLPT2-MP2 gap was already small, and now further shrinks to just 0.2 kcal/mol.

## CONCLUSIONS

The research question was whether the evaluation of the nonlocal correlation term in a basis of Kohn-Sham orbitals (i.e., a fifth-rung DFT functional) has any advantages over using standard HF-MP2 correlation (i.e., a composite method constructed from hybrid DFT and standard MP2). The answer to the question is definitely affirmative. However, while the use of GLPT2 rather than HFMP2 nonlocal correlation reduces WTMAD2 (weighted mean absolute deviation) for the GMTKN55 benchmark suite by about one-third, such a difference is smaller than one might be led to expect. Part of the improvement appears to stem from the greater resilience of KS orbitals to spin contamination, which causes significant improvements for radical reactions (e.g., in BH76 and RSE43). However, even when excluding the latter, or using ROMP2 instead of UMP2, GLPT2-based double hybrids still hold a significant edge over their composite WFT-DFT counterparts.

## ACKNOWLEDGMENTS

This research was supported by the Israel Science Foundation (grant 1969/20) and by the Minerva Foundation (grant 2020/05). G.S. acknowledges a fellowship from the Feinberg Graduate School of the Weizmann Institute.